\documentstyle[aps,psfig,preprint]{revtex}

\begin{document}

\title{{\bf Polarization coupling and pattern selection in a type-II optical
parametric oscillator}}

\author{{\it Gonzalo Iz\'us\cite{Gonzalo}, Maxi San Miguel and Daniel
Walgraef \cite{Daniel}}}
\address{Instituto Mediterr\'aneo de Estudios Avanzados, IMEDEA (CSIC-UIB)
\cite{www},\\
Universitat de les Illes Balears, E-07071 Palma de Mallorca, Spain.}

\maketitle

\vskip 1.5cm

\tightenlines

\begin{abstract}
We study the role of a direct intracavity polarization coupling in
the dynamics of transverse pattern formation in type-II optical
parametric oscillators. Transverse intensity patterns are
predicted from a stability analysis, numerically observed, and
described in terms of amplitude equations. Standing wave intensity
patterns for the two polarization components of the field arise
from the nonlinear competition between two concentric rings of
unstable modes in the far field. Close to threshold a wavelength
is selected leading to standing waves with the same wavelength for
the two polarization components. Far from threshold the
competition stabilizes patterns in which two different wavelengths
coexist.
\end{abstract}

\vskip 1.5cm
\newpage

\section{INTRODUCTION}

Pattern formation is an ubiquitous manifestation of nonlinearity
\cite{cros93,Daniel96} which presents specially interesting
features in nonlinear optical systems
\cite{pinos,arecchi,Lugiatoreview}. The search of transverse
structures in nonlinear optical systems is actively pursued for
several reasons which include their possible application in
all-optical signal processing and the investigation of macroscopic
manifestations of quantum phenomena. These structures are the
result of the interaction of nonlinearity and diffraction in
transverse spatially extended devices such as nonlinear optical
cavities of large Fresnel number. Among the nonlinear systems
analyzed, optical parametric oscillators (OPO's) have received a
lot of attention from the theoretical viewpoint. Available results
include the analysis of pattern formation
\cite{oppo94,valcarcel96,longhi,taki,saffman}, noise sustained
structures \cite{marcoNSS} domain walls \cite{tril97,long97,op1}
and localized structures \cite{stal98,oppo99,leberre,emilio}. A
growing interest in these transverse structures in OPOs arise also
from the study of quantum spatial correlations present in these
patterns\cite{Lugiatoreview,lu-ca,gatti,zambrini2}. Transverse
patterns in OPO have been recently observed \cite{fabr99}.

In an OPO two first harmonic (FH) fields (signal and idler) are
generated inside the crystal by parametric down conversion of the
external pump field. In type-I OPO signal and idler fields have
the same state of linear polarization. In type-II OPO they are
orthogonally  polarized. This polarization degree of freedom can
be used for a nonlinear construction of new states of the emitted
light. For example, by means of a direct polarization coupling
produced by an intracavity  quarter wave-plate ($\lambda/4$-plate)
it is possible to produce \cite{Mason98,Fabre00} states in which
signal and idler are degenerate in frequency and phase locked. A
general question that we address in this paper is the effect of
this type of direct polarization coupling in the problem of
transverse pattern formation in type-II OPO.

When considering transverse spatial degrees of freedom in type-II
OPO without a direct polarization coupling there are two different
regimes. In one of them, characterized by a positive effective
detuning homogeneous solutions are selected at threshold with an
arbitrary relative phase between signal and idler fields. For
effective negative detuning a finite wave number is selected at
threshold  and a phase pattern  --traveling wave (TW)-- is
asymptotically selected for each FH field, while the intensity
remains homogeneous \cite{longhi96,Nos99}. The effects of direct
polarization coupling between signal and idler in type-II OPO for
a positive effective detuning was discussed in refs.
\cite{Nos00,Nos01}: Spatial domains of equivalent, but different,
self-phase locked homogeneous solutions appear. They are separated
by phase polarization domain walls. These walls are generally
Bloch walls whose motion leads to complex spatiotemporal states.
In this paper we consider the situation of negative effective
detuning in order to characterize the way in which polarization
coupling modifies the process of pattern formation. In particular
one expects that the coupling between signal and idler can
generate standing waves (SW큦) --i.e, stripe intensity patterns--
from the TW큦 that exist for signal and idler when there is no
direct polarization coupling.

For a type-II OPO with an intracavity $\lambda/4$-plate and for
FH's negative detunings we predict, and numerically confirm, that
there is a threshold of pattern formation above which the FH's far
fields exhibit during the transient dynamics two concentric rings
of growing unstable modes. Therefore, the coupling does not only
lead from TW큦 to SW큦, but it introduces two different
wavelengths, giving rise to an interesting problem of wavelength
competition and pattern selection. The values of the different
wavelengths are controlled by the $\lambda/4$-plate polarization
coupling. In a ``symmetric case" (i.e, all the dynamical
parameters for signal and idler are equal) we show that the two
wavelengths coexist for long times. Real and imaginary parts of
each FH field present in this case SW patterns with different
wavelength which consist in domains of stripe patterns. For
asymmetric FH's coefficients and near threshold, we show that one
of the wavelengths dominates and intensity  stripe patterns
-standing waves- with the same wavelength emerge for signal and
idler. Far from threshold the dynamics is similar to the symmetric
case. This transition from wavelength coexistence  to dominance of
one of the two competing wavelengths is described by means of an
amplitude equation analysis. The amplitude equations give a full
description of the observed dynamics.

The paper is organized as follows. Section 2 reviews the mean
field equations for this system. In section 3 we analyze the
threshold for pattern formation. The instability is characterized
in terms of the eigenfunctions associated with the critical modes.
In section 4 we discuss numerically the dynamics of transverse
pattern.  In section 5 we derive amplitude equations for the
critical modes, which give a clear interpretation of the observed
dynamics. We summarize our main conclusions in section 6.

\section{MEAN FIELD EQUATIONS}

We consider an optical parametric oscillator that consists of a
ring optical cavity filled with a birefringent, nonlinear
quadratic medium and externally pumped by a uniform  laser beam. A
direct polarization coupling between the FH fields, that take into
account the effect  of an intracavity quarter wave-plate is also
included in the model. This wave plate provides a polarization
mixing between  the signal $A_x$ and idler $A_y$ fields
\cite{Mason98,Fabre00}. The signal and the idler can be either
frequency degenerate or non-degenerate, depending on the frequency
selection rules imposed by the combined effects of the parametric
down-conversion, the cavity resonances and phase-matching
\cite{falk71,byer91,fabr93}, but they are always polarization
non-degenerate (type-II interaction). In the mean field
approximation, and considering the paraxial and the single
longitudinal mode approximation for all the fields, the equations
describing the time evolution for the linear polarization
components of the second harmonic ($B_{x,y}(x,y,t)$) (SH) and the
first harmonic ($A_{x,y}(x,y,t)$) (FH) slowly varying envelopes of
the electric fields, in a type-II, phase-matched OPO are
\cite{Fabre00,Nos01}:
\begin{eqnarray}
\partial_t B_x & = & \gamma_x' [-(1+i \Delta_x')
B_x + i \alpha_x' \nabla^2 B_x + 2i K_0 A_x A_y + E_0+c' B_y] \nonumber \\
\partial_t B_y & = & \gamma_y' [-(1+i \Delta_y') B_y + i \alpha_y' \nabla^2
B_y
-c'^* B_x] \nonumber \\
\partial_t A_x & = & \gamma_x [-(1+i \Delta_x) A_x + i \alpha_x \nabla^2 A_x
+
i K_0 A_y^* B_x +c A_y] \nonumber \\
\partial_t A_y & = & \gamma_y [-(1+i \Delta_y) A_y + i \alpha_y \nabla^2 A_y
+
i K_0 A_x^* B_x -c^* A_x]
\label{master}
\end{eqnarray}
where with no loss of generality we take $A_x,B_x$ as ordinary
polarized beams and $A_y,B_y$ as extraordinary polarized
\cite{Nos99}. The coefficients $\gamma_{x,y}, \gamma_{x,y}'$
(cavity decay rates), $\Delta_{x,y}, \Delta_{x,y}'$  (cavity
detunings) and $\alpha_{x,y},\alpha_{x,y}'$ (diffraction
coefficients) are defined as in refs. \cite{oppo94,Nos00}; due to
the birefringence of the nonlinear crystal all the diffraction
coefficients can be slightly different, even when the signal and
idler are frequency degenerate. Other parameters are the
nonlinearity $K_0$ and the injected pump $E_0$ (bifurcation
parameter) that, for the sake of simplicity, we take it real and
polarized along the same direction than the phase-matched
component of the second harmonic field $B_x$. Hence, $B_y$ neither
is pumped nor is nonlinearly coupled with other components of the
field. Under these conditions $B_y$ does not influence the
dynamics although it is linearly coupled to $B_x$. In
eqs.(\ref{master}) the direct polarization coupling constants
($c,c'$) account for the effects produced by the $\lambda/4$-plate
and they are related to the phase mismatch and the axes of the
$\lambda/4$-plate  by:
\begin{equation}
c \sim  \sin(2 \phi) \, \exp(i \xi)
\label{coupling}
\end{equation}
where $\phi$ is the angle between the $\lambda/4$-plate's fast
axis and the principal axis of the crystal. The phase $\xi$ is the
round-trip phase shift between the signal and idler at frequency
degeneracy (or between $B_x$ and $B_y$ for $c'$). The coupling
strength $|c|$ depends on a number of factors, including mode
matching and Poynting vector walk-off \cite{Mason98}. Here we
assume perfect phase-matching and propagation along the optical
axis (i.e. we neglect spatial walk-off). We focus on  the effects
produced by the rotation angle $\phi$ which  is an important
experimental parameter to control the effects described below. We
note that other forms of similar linear coupling terms between
signal and idler considered in \cite{Nos00,Nos01} are associated
with a birefringent and/or dichroic cavity mirror in  type-II OPO.

The linear coupling  $c$ breaks the phase invariance, that
eqs.(\ref{master}) have for $c=0$, under changes of the relative
phase of the FH fields: $[A_x,A_y] \rightarrow  [\exp(i \varphi)
\, A_x,\exp(-i \varphi) \, A_y]$. However, the phase of the signal
and idler fields can be adjusted in order to include the phase of
$c$ in the FH fields; i.e, given $c=|c| \, \exp(i \xi)$, the
transformation: $[A_x,A_y] \rightarrow [\hat{A}_x,\hat{A}_y] =
[\exp(-i \xi/2) \, A_x,\exp(i \xi/2) \, A_y]$ leaves
eqs.(\ref{master}) unchanged except for the replacement $c
\rightarrow |c|$. For the sake of clarity we will then present our
main results for $c$ real. The generalization for $c$ complex is
trivial using the phase transformation of the FH fields just
described.

\section{LINEAR STABILITY ANALYSIS}

\subsection{Threshold analysis}

In this section we present the linear stability analysis of the
steady-state solution of eqs.(\ref{master}) corresponding to the
OPO operating below the threshold of signal generation. This
trivial uniform steady state (off-state) is:
\begin{eqnarray}
A_{x} & = & A_{y}= 0  \nonumber \\
B_x & =& (1+i\Delta_y') \, E_0/[1-\Delta_x'
\Delta_y'-|c'|^2+i(\Delta_x'+\Delta_y')] \nonumber\\
B_y & = & -c'^* \, B_x/(1+ i \, \Delta_y')
\label{trivial}
\end{eqnarray}
The threshold for transverse pattern formation is determined by
linearizing eqs.(\ref{master}) around this solution and looking
for instabilities. The steady state becomes unstable only along
the directions of the FH components ($A_x,A_y$) of the
eigenvectors, and thus the analysis reduces to the study of two
linearly-coupled complex equations. Because of the complex nature
of the field variables, it is convenient to consider the real and
the imaginary parts of these equations for each FH field. The most
general solution of the perturbations  is hence given by a linear
superposition of terms of the form: $[Re (A_{x,y}), Im(A_{x,y}) ]
\sim \exp[i \vec q \, \vec r+\lambda(\vec q) t] $ where $\lambda
(\vec q)$ is the growth rate of the perturbations and $\vec q$ is
its  transverse wave vector.

For $c=c'=0$, the linear stability analysis shows that the trivial
solution is stable for $|F|<|F_c|$, where $F$ is a normalized pump
intensity
\begin {equation}
F=K_0 \, E_0/(1+i\Delta_x')
\label{norpump1}
\end{equation}
For $\tilde \Delta=\gamma_x \Delta_x + \gamma_y \Delta_y
>0$ the most unstable mode corresponds to an homogeneous solution
$q_0=0$. In this paper we focus on the case $\tilde \Delta < 0$
for which the unstable modes at threshold ($|F_c|=1$) correspond
to transverse traveling waves $A_x, A_y^* \simeq \exp [i \vec q \,
\vec r + \lambda (\vec q ) \, t ]$, whose two-dimensional
wave-vector $\vec q$ lies on a circle centered at $0$ with radius
$q_0=\sqrt{-\tilde \Delta/\tilde \alpha}$, where  $\tilde
\alpha=\gamma_x \alpha_x + \gamma_y \alpha_y $. For $|F|>|F_c|$
and $\tilde \Delta<0$ any $\vec q$ mode on the circle, and the
opposite mode for the orthogonal component of the field, can be
selected at threshold by means of spontaneous symmetry breaking.
Hence a phase pattern appears above threshold for $A_x$ and $A_y$
-traveling waves-, with opposite wavevector, while the intensity
remains homogeneous in both polarizations \cite{Nos99,longhi96}.
It should be noted that when idler and signal fields are
degenerate both in frequency and polarization --i.e, type-I
DOPO--, eqs.(\ref{master}) must be solved with the further
condition $A_x=A_y$. In this case standing waves states are
selected at threshold\cite{oppo94}. The linear coupling of $A_x$
with $A_y$ considered in this paper is expected to produce
standing waves for each polarization component also in type-II
OPO.

For $c \neq 0$ the threshold for pattern formation remains at
$|F_c|=1$, but a main difference is that now the wavevectors of
the most unstable modes lie on two concentric circles of different
radius. This gives rise to a wavenumber competition in the process
of pattern formation, as shown below. Closed expressions are hard
to obtain in the general case, but for $\gamma_x=\gamma_y=\gamma$
the two eigenvalues of the linearized equations that characterize
the instability are
\begin{equation}
\lambda_{1,2}  =  \gamma \, \left[ -1 + \frac{1}{2} \,
\sqrt{4 \, | F|^2-4 \, | c|^2 - 2 \, (\Theta_x^2+\Theta_y^2) \pm  2
\Upsilon}  \right]
\label{eigen0}
\end{equation}
where we have defined:
\begin{eqnarray}
\Theta_j & = & \Delta_j+\alpha_j \, q^2 \,\,\,\,\,\,\, \mbox{($j=x,y$)}
\nonumber \\
\Upsilon  & = & \sqrt{-4 \, |F|^2 \, (\Theta_x-\Theta_y)^2 +
4 \,| c|^2 \,(\Theta_x+\Theta_y)^2+(\Theta_x^2-\Theta_y^2)^2}
\end{eqnarray}
and we have introduced the normalized pump amplitude:
\begin{equation}
F=(1+i\Delta_y') \, K_0 \, E_0/[1-\Delta_x'
\Delta_y'-|c'|^2+i(\Delta_x'+\Delta_y')]
\label{norpump2}
\end{equation}
which coincides with $F$ given by eq.(\ref{norpump1}) for $c'=0$.
Two other eigenvalues remain always negative.

To avoid cumbersome expressions, analytical results are derived in
this section for the particular case $\gamma_x=\gamma_y=\gamma$,
$\alpha_x=\alpha_y= \alpha$ and $\Delta_x=\Delta_y=\Delta$ ($<0$).
In this case, the eigenvalues $\lambda_{1,2}(\vec q)$ become:
\begin{equation}
\lambda_{1,2}  =  \gamma \, \left[ -1 +  \sqrt{| F|^2-(\Delta+\alpha \,
|\vec q|^2 \pm c)^2}  \right]
\label{eigenval}
\end{equation}
where the plus (minus) sign corresponds to $\lambda_1$
($\lambda_2$). From eq.(\ref{eigenval}) we get the threshold of
instability for perturbations with an arbitrary wavevector $\vec
q$:
\begin{equation}
|F_{1,2}(c) |^2 =1+ \left[ \Delta+ \alpha \, |\vec q|^2 \pm c
\right]^2 \label{threshold}
\end{equation}
where the plus (minus) sign corresponds to $F_1$ ($F_2$).
Therefore, for $c < -\Delta$ the instability takes place at the
critical threshold $|F_c|=1$ and the  unstable modes at threshold
correspond to transverse traveling waves whose two-dimensional
real wave-vector $\vec q$ lies on either of two concentric
circles, centered at $0$ with radius $q_{1,2}$:
\begin{equation}
q_{1,2}^2=|\vec q_{1,2}|^2= \frac{-\Delta \mp c }{\alpha}
\label{radio}
\end{equation}
In figure 1 we show the instability threshold for perturbations of
different wavenumbers. The threshold $F_c=1$ is the same for $c =
0$ and $c \neq 0$, but for $c \neq 0$ the instability takes place
at two different wavevectors of modulus $q_{1,2}$ indicated in the
figure. Homogeneous perturbations ($\vec q=0$) have a larger
instability threshold. The homogeneous phase locked solutions
associated with this threshold are discussed in the appendix. We
will not consider here the case $c > -\Delta >0$ for which at
$F_c=1$ only the mode $q_2$ becomes unstable. In this case the
mode $q_1=0$ becomes unstable for larger values of the pump.

The values of the most unstable wavenumbers $q_1$ and $q_2$ depend
on the absolute value of $c$ as follows from eq.(\ref{radio})
(figure 2). In particular, $q_{1,2}^2$ coincide with $q_0^2$ for
$c=0$ and vary linearly with $|c|$. The polarization coupling
splits the circle of unstable modes for $c=0$ in two circles. The
value of $c$ controls the magnitude of the split. We remark that
for $c \sim | \Delta |$, $|\vec q_1| \sim 0$ and patterns with
very large wavelength can be expected. For $|F|>|F_c|$ there is a
band of unstable modes associated with each eigenvalue. In figure
3 we show the real part of the eigenvalues $\lambda_{1,2}$ as a
function of $|\vec q|$ for the critical case ($|F|=|F_c|=1$) and
for one case above threshold. All the modes with
$Re(\lambda_{1,2})>0$ are linearly unstable. The wavevectors with
modulus $|\vec q_1|$ or $|\vec q_2|$ have the same maximum growth
rate: $\lambda_{1,2}(\vec q_{1,2})= \gamma \, (-1+|F|)$.

When the damping, detuning or diffraction parameters for signal
and idler are different the competing modes of wavenumbers $q_1$
and $q_2$ have different thresholds. It follows from the numerical
analysis of eq.(\ref{eigen0}) that the smaller wavenumber $q_1$
becomes first unstable at $|F_c|=1$. In figure 4a we show the
growth rate for perturbations of the trivial state as a function
of $|\vec q|$ for a case in which $|F|=1.0019$. The unstable mode
of smaller wavenumber $q_1$ dominates in the linear regime. The
difference in growth rates for $q_1$ and $q_2$ depend on the
pumping level. For larger values of the pump, the growth rates of
both unstable modes are of the same order, being the growth rate
of $q_1$ larger than the one of $q_2$ (see fig. 4.b). In the next
section we show that this fact deeply affects the nonlinear mode
competition dynamics of the system.

We finally note that for asymmetric FH's coefficients, the
critical modes have in general real eigenvalues (i.e.
$Im(\lambda_{1,2}(\vec q_{1,2}))=0$). However, for very small
values of $c$, i.e., when $|\vec q_1| \sim |\vec q_2|$ there is a
small interval of values of the external pump $F$, which includes
the critical value $F_c=1$, where $Im(\lambda_{1,2}(\vec q_{1,2}))
\neq 0$. In this limit of $c \rightarrow 0$ the instability
becomes convective, similarly to a situation considered in
ref.\cite{convec}.

\subsection{Critical modes}

Next we consider some features of the early time dynamics of
pattern formation that can be understood in terms of the
eigenvectors corresponding to the eigenvalues of the linearized
problem discussed above. First, we introduce the far field  as the
Fourier transform of the near field, where the near field is the
transverse field configuration at the input/output cavity mirror.
The far field components $\widetilde{A}_{\vec q} (t)$ of $A_x$
(for example) are defined by:
\begin{equation}
A_x(\vec r,t) = \frac{1}{2 \pi} \int_{- \infty}^{\infty}
\widetilde{A}_{\vec q}(t) \, \exp(i \vec q \, \vec r) \, dq_x  \,
dq_y
\end{equation}

In figure 5 we show numerical results \cite{numerical} for a
typical transverse profile of the $A_x$ and $A_y$ FH fields at an
early time after the pump is increased beyond its threshold value.
These results correspond to the case of symmetric coefficients in
which two competing wavenumbers have the same growth rate. In
Fig.5a we show the near field of the signal intensity pattern and
its far field. The two concentric rings of the far field
correspond to unstable wavectors $\vec q$ with arbitrary
orientation and wavenumber around $q_1$ (inner ring) and $q_2$
(outer ring). The near field is the result of the interference
among all the unstable modes of both rings in the far field.
However, the interference takes place in such a way that the real
part of $A_x$ is associated with the unstable modes of the outer
ring, while its imaginary part is associated with the unstable
modes of the inner ring. In addition there is a high correlation
between the transverse structures observed in the signal $A_x$ and
idler $A_y$ fields. This is illustrated in Figs. 5b, 5c and 5d. In
Figs. 5b and 5c we show the near field of the real and imaginary
parts of $A_x$ and $A_y$. It is observed that the real and
imaginary parts of the field $A_x$ support transverse patterns
with different wavelength. The same fact is observed in the idler
field $A_y$, but real and imaginary part have a different
wavelength than for $A_x$. In fact we observe that $Re(A_x)\simeq
Im(A_y)$, while $Im(A_x)\simeq Re(A_y)$. In Fig. 5d we show the
far field of $Re(A_x)$ and $Im(A_x)$. This gives evidence of the
different wavenumber associated with $Re(A_x)$ and $Im(A_x)$. The
two rings in the far fields of $Re(A_x)$ and $Im(A_x)$ correspond,
respectively, to the outer and inner rings of Fig. 5a.

These numerical facts can be explained in terms of the
eigenvectors $\Lambda_{1,2}(\vec q)$ associated with the
eigenvalues $\lambda_{1,2}$ introduced in eq.(\ref{eigenval}).
They can be written as
\begin{eqnarray}
\left[ Re (A_x), Im (A_x),Re (A_y), Im (A_y) \right]^T  =
\Lambda_1(\vec q) & = & C_1 \, \left[\kappa_1,1,1,-\kappa_1
\right]^T \, \exp(i \vec q \, \vec r)
\nonumber \\
\Lambda_2(\vec q) &= & C_2 \, \left[ 1,-\kappa_2,\kappa_2,1
\right]^T \, \exp(i \vec q \, \vec r) \label{eigenf}
\end{eqnarray}
where we have introduced:
\begin{eqnarray}
\kappa_1 & = & \frac{-\lambda_1+F-1}{\Delta+\alpha \, |\vec q|^2 + c}
\nonumber \\
\kappa_2 & = & \frac{-\lambda_2+F-1}{\Delta+\alpha \, |\vec q|^2 - c}
\end{eqnarray}
and the normalization constants $ C_j^{-1}=\sqrt{2 \,
(1+|\kappa_j|^2)}, \, \, j=1,2$. For simplicity we take here $F$
to be real (i.e. resonant pump field). Exactly at threshold,
$\kappa_{1,2}$ vanish for the corresponding critical wave-vector:
$\kappa_{1,2}(\vec q_{1,2})=0$. The general dependence of
$\kappa_{1,2}$ on the wavenumber at threshold is shown in Fig. 6.
At threshold $F=F_c=1$ and the eigenvectors $\Lambda_{1,2}(\vec
q)$ are damped for any $\vec q$ except $\Lambda_{1}(\vec q=\vec
q_1)$ and $\Lambda_{2}(\vec q=\vec q_2)$:
\begin{eqnarray}
\Lambda_1(\vec q_1) & = & \frac{1}{\sqrt{2}} \,
\left[ 0,1,1,0 \right]^T \, \exp(i \vec q_1 \, \vec r) \nonumber \\
\Lambda_2(\vec q_2) &= &\frac{1}{\sqrt{2}} \, \left[ 1, 0 , 0, 1
\right]^T \, \exp(i \vec q_2 \, \vec r) \label{eigenf2}
\end{eqnarray}
which are marginal (zero growth rate) and define the direction in
the functional space along which the instability takes place. The
form of the eigenvectors (\ref{eigenf2}) explain our numerical
finding (Fig.5) that at short times, when the linear approximation
to the dynamics remains valid, the components of the FH's fields
($Re(A_x), Im(A_y)$) and ($Im(A_x),Re(A_y)$) only sustain patterns
with wavevector of modulus $|\vec q_2|$ or $|\vec q_1|$
respectively. As we discuss in the next section, this gives rise
to pattern formation with competing wavelengths that can be rather
different.

There is an interesting symmetry in the far field components of
the first harmonics fields. We focus in the $A_x$ field, but the
discussion is also valid for $A_y$. Our numerical results
indicate, as shown for example in Fig. 7,  that the real
(imaginary) part of the far field component $\widetilde{A}_{\vec
q} (t)$ is an odd function for $\vec q= \vec q_1$ and an even
function for $\vec q= \vec q_2$ (even function for $\vec q= \vec
q_1$ and an odd function for $\vec q= \vec q_2$). This property
implies that
\begin{eqnarray}
\widetilde{A}_{-\vec q_1} & = & -  \widetilde{A}_{\vec q_1}^*   \nonumber \\
\widetilde{A}_{-\vec q_2} & = &  \widetilde{A}_{\vec q_2}^*
\label{amplitude}
\end{eqnarray}
Given that in the linear regime at threshold only excitations with
wavectors $\vec q_2$ or $\vec q_1$ contribute to the resulting
structures in the real-valued fields $Re(A_x)$ or $Im(A_x)$,
respectively, we find that the basic excitations for $Re(A_x)$ and
$Im(A_x)$ are standing waves of the form
\begin{eqnarray}
Re(A_x) & \sim & \widetilde{A}_{\vec q_2} \, \exp(i \vec q_2 \vec
r)+
\widetilde{A}_{- \vec q_2} \, \exp(-i \vec q_2 \vec r) \nonumber \\
i Im(A_x) & \sim & \widetilde{A}_{\vec q_1} \, \exp(i \vec q_1
\vec r)+ \widetilde{A}_{- \vec q_1}  \, \exp(-i \vec q_1 \vec r)
\label{superp}
\end{eqnarray}
These standing waves are the interference between two opposite
modes of the same ring of the far field which satisfy
eq.(\ref{amplitude}). Both modes have the same amplitude but the
global phase of the superposition of the two modes is different in
each ring.

It is interesting to note that the circularly polarized components
$A_\pm$ of the vectorial FH field give a natural description of
the instability. We have already discussed that, at threshold
$Re(A_x)\simeq Im(A_y)$, and $Im(A_x)\simeq Re(A_y)$, so that
$A_+=(A_x+iA_y)/\sqrt{2}=i\sqrt{2} \, Im(A_x)$ and
$A_-=(A_x-iA_y)/\sqrt{2}=\sqrt{2} \, Re(A_x)$. Therefore, it
follows from eq.(\ref{superp}) that the instability for  $A_+$
($A_-$) takes place at $q_1$ ($q_2$). The two circularly polarized
components will emerge at the instability as standing waves
intensity patterns of different wavenumber. In fact, for symmetric
coefficients and in the linearized version of eqs. (1) around the
trivial solution (\ref{trivial}), $A_+$ and $A_-$ are decoupled.
They are nonlinearly coupled through the pump field $B_x$.

Our above discussion is for real values of the parameter $c$. When
$c$ is complex ($\xi \neq 0$) both the real and imaginary part of
$A_x$ and $A_y$ have contributions of unstable modes of wavenumber
$q_1$ and $q_2$. Therefore we observe  transverse patterns with
competing wavelengths in the real and the imaginary parts of
$\hat{A}_{x,y}$. However, by changing the global phase of the FH
fields, the problem can be considered in terms of $c$ real, as
previously pointed out.

\section{TRANSVERSE INTENSITY PATTERNS}

In this section we give a numerical description \cite{numerical}
of the patterns that are asymptotically selected after a regime of
nonlinear competition among the unstable modes of wavenumber $q_1$
and $q_2$ . A theoretical justification of these results is given
in the next section in terms of an amplitude equation. We consider
separately the cases of symmetric and nonsymmetric coefficients
for the FH fields.

\subsection{Symmetric coefficients}

When $\gamma_x=\gamma_y=\gamma$, $\alpha_x=\alpha_y= \alpha$ and
$\Delta_x=\Delta_y=\Delta$ the growth rates of the most unstable
modes in the two circles of radius $q_1$ and $q_2$ are  equal and
the instability takes place at both circles simultaneously (see
Fig. 3). In this case the nonlinear competition keeps wavevectors
in both circles in the far field excited for long times. The real
and imaginary part of the FH fields show patterns with different
wavelengths, as discussed in section 3. To illustrate the
dynamical evolution we show in Figs. 8 and 9 snapshots of the
pattern configuration at two different times. The two rings in the
far field persist at long times. The pattern which appears in the
long time dynamics presents domains in which real and imaginary
parts of each of the FH near fields show standing waves of
arbitrary orientation and of different wavelength for the real and
imaginary parts. The arbitrary orientation of these standing waves
comes from the spontaneous choice of two opposite wavevectors in
the corresponding ring of the far field. Thus, the general picture
is that there is long time competition among standing waves of
different orientations and two different wavenumbers.

A limiting situation is the case in which the inner ring collapse
to the point $\vec q=0$. This situation takes place when
$c=|\Delta_x|$. In this case, the growth of uniform domains occurs
in one of the components of the FH vector field while the other
component sustains standing waves of local arbitrary orientation
as shown in figure 10. This structure represents a case in which
patterns and uniform domains coexist in the same complex field as
an effect induced by the direct polarization coupling. This
situation takes place near signal resonance for very small values
of $c$.

\subsection{Nonsymmetric coefficients}

When the damping, diffraction or detuning coefficients of signal
and idler are different, the nonlinear mode competition depends
very much on how far above threshold the OPO is pumped. Near
threshold the relative difference in the magnitude of the growth
rate of the unstable modes on the circles of wavenumbers $q_1$ and
$q_2$ is important, as follows from Fig. 4.a. This fact produces a
strong change in the dynamics of the system. In Figs. 11, 12 and
13 we show snapshots of configurations at different times of the
dynamical evolution. After a transient linear regime, discussed in
section 3 and which is represented here by Fig. 11, there is a
nonlinear competition between the two circles of unstable modes at
intermediate times. This is shown in Fig. 12 where the far field
of $Re(A_x)$ and $Im(A_y)$ are seen to have competing
contributions from the two circles. At late times the inner circle
wins the competition and the final pattern is a standing wave of
wavenumber $q_1$ both for $A_x$ and $A_y$. This nonlinear
wavenumber selection can be traced back to the behavior of the
growth rate, as shown in the next section. For long times there is
also a spontaneous breaking of the rotational symmetry and a
standing wave in an arbitrary direction is selected, as shown in
Fig.13. The real and imaginary part of the FH fields show in this
case patterns with the same wavelength. The resulting structures
originate in the interference between two arbitrary -but opposite-
wavevectors of the inner circle. The resulting stripe intensity
pattern is similar to that predicted for type-I DOPO in the sense
that it is the interference between two opposite traveling waves
\cite{oppo94}. However, physically, the energy and momentum
conservation in the parametric down conversion of pump photons
only implies off-axis emission of idler and signal photons along
two opposite directions without interference between them because
they have orthogonal polarization. Due to polarization coupling,
pure traveling wave are not solutions  of eqs.(\ref{master}). The
$\lambda/4$-plate provides a mixing of polarization that allows
that photons of the same FH field interfere  producing a standing
wave. This phenomenon is the same that occurs in  resonantly
coupled complex Ginzburg-Landau equations \cite{prl96}: the linear
(polarization) coupling allows the formation of standing waves as
the result of the interaction between opposite modes in the far
field --in this case, the inner ring--.

Far from threshold, the growth rates of the unstable modes $q_1$
and $q_2$ are of the same order (see Fig. 4.b) and the dynamics of
the system in this regime is equivalent to the symmetric case, as
we prove in the next section. Therefore, in the asymmetric case
the intensity of the external pump can be used to stabilize both
rings of the far field which have a competing coexistence for long
times and far from threshold. In figure 14 we show a typical long
time state for this regime.

\section{AMPLITUDE EQUATIONS}

Close to the instability threshold, and using the general methods
of nonlinear dynamics and pattern formation theory one may derive
amplitude equations for the patterns described numerically in the
previous section. As it is now well known, these equations are
able to describe pattern evolution, selection and stability. In
particular, if, for simplicity, one considers only the critical
modes identified in section 3, field variables may by expressed
as:
\begin{eqnarray}
& & \left[ Re (A_x), Im (A_x), Re (A_y), Im (A_y) \right]^T   =
A_1 \,    \left[ 0,1,1,0 \right]^T \, \exp(i \vec q_1 \vec r) +
A_{-1} \,\left[ 0,1,1,0 \right]^T \, \exp(-i \vec q_1 \vec r)
\nonumber \\
& + & A_{2} \,\left[ 1,0,0,1 \right]^T \, \exp(i \vec q_2 \, \vec
r) +  A_{-2} \,\left[ 1,0,0,1 \right]^T \, \exp(-i \vec q_2 \,
\vec r) \label{fields}
\end{eqnarray}
where $A_i(t)$ are slowly evolving amplitudes associated with the
modes with wavevector $\vec q_i$. Given the relations
(\ref{amplitude}) we have chosen here to define the amplitudes
$A_{\pm 1}$, $A_{\pm 2}$ as $\widetilde{A}_{\pm \vec q_1} = i
A_{\pm 1}$ and $\widetilde{A}_{\pm \vec q_2} = A_{\pm2}$

Standard analysis \cite{Daniel96} leads to the following evolution
equations for the symmetric case ($\Theta_x=\Theta_y$ and
$c'=\Delta_x'=0$ for simplicity)
\begin{eqnarray}
\partial_t A_{\pm 1} & = & \gamma (\vert F \vert - 1) A_{\pm 1} +
\frac{K_0}{2} [A_{\pm 1} \, (\widetilde{B}_x+\widetilde{B}_x^*)_0
+A_{\mp 1} \, (\widetilde{B}_x+\widetilde{B}_x^*)_{\pm 2 \vec q_1}
+  i A_2 \, (\widetilde{B}_x-\widetilde{B}_x^*)_{\pm \vec q_1
-\vec q_2}
\nonumber\\
& + & i A_{-2} \, (\widetilde{B}_x-\widetilde{B}_x^*)_{\pm \vec q_1 +\vec
q_2}]   \nonumber\\
\partial_t A_{\pm 2} & = & \gamma (\vert F \vert - 1) A_{\pm 2} +
\frac{K_0}{2}
[A_{\pm 2} \, (\widetilde{B}_x+\widetilde{B}_x^*)_0 +A_{\mp 2}
\, (\widetilde{B}_x+\widetilde{B}_x^*)_{\pm 2 \vec q_2}
+   i A_1 \, (\widetilde{B}_x-\widetilde{B}_x^*)_{\pm \vec q_2 -\vec q_1}
\nonumber\\
& +i & A_{-1}\,(\widetilde{B}_x-\widetilde{B}_x^*)_{\pm \vec q_1 +\vec q_2}
]
\label{modemaster}
\end{eqnarray}
where
\begin{equation}
\partial_t (\widetilde{B}_x)_{\vec q} = -(1+i \alpha' q^2) \, (\widetilde{B}_x)_{\vec q}
+2 i K_0 (\widetilde{A_x A_y})_{\vec q}  \label{modemaster2}
\end{equation}
$\alpha' =\alpha_x'$ and $(\widetilde{...})_{\vec q}$ indicates
Fourier transform. Using eqs.(\ref{modemaster2}) and time scales
separation near threshold ($\vert F \vert -1 \ll 1$) one may write
$(\widetilde{B}_x)_j=B_j$ as an expansion in the amplitudes $A_i$.
One has, at lowest order:

\begin{eqnarray}
B_0 & = & -4 K_0 \, (A_1 A_{-1}+A_2 A_{-2})  \nonumber\\
B_{\pm 2 \vec q_i}  & = &
-\frac{2 K_0 A_{\pm i} A_{\pm i}}{1+4 i \alpha' \, q_i^2 }\nonumber\\
B_{\pm \vec q_2 \pm  \vec q_1}  & \simeq & 0
\end{eqnarray}

This leads to:

\begin{eqnarray}
\partial_t A_{\pm 1} & = & \gamma (\vert F \vert - 1) A_{\pm 1} -4 K_0^2
A_{\pm 1}^2 A_{\mp 1} -4 K_0^2 A_{\pm 1} A_{2}A_{-2}-4 \eta_1K_0^2
A_{\mp 1} A_{\pm 1}^2 \nonumber\\
\partial_t A_{\pm 2} & = & \gamma (\vert F \vert - 1) A_{\pm 2} -4 K_0^2
A_{\pm 2}^2 A_{\mp 2} -4 K_0^2 A_{\pm 2} A_{1}A_{-1}-4 \eta_2K_0^2
A_{\mp 2} A_{\pm 2}^2
\end{eqnarray}
where $\eta_{1,2}^{-1} = 2  \, [1+(4 \alpha' \, q_{1,2}^2)^2]$

Since  $A_{-1}=A_1^*, \, A_{-2}=A_2^*$, as follows from
(\ref{amplitude}) and the definition of the amplitudes $A_{\pm
1}$, $A_{\pm 2}$ after eq. (\ref{fields}), one may finally write:
\begin{eqnarray}
\partial_t A_{ 1} & = & \gamma (\vert F \vert - 1) A_{ 1} -4 K_0^2 \, A_ 1
\, \lbrack (1+\eta_1)|A_1|^2+|A_2|^2\rbrack \nonumber\\
\partial_t A_{2} & = & \gamma (\vert F \vert - 1) A_{2} -4 K_0^2 \, A_ 2 \,
\lbrack (1+\eta_2)|A_2|^2+|A_1|^2\rbrack
\label{finalmaster}
\end{eqnarray}
These equations describe a weak competition between the amplitudes
$A_1$ and $A_2$, so that the stable steady states of the system
are given by:
\begin{eqnarray}
|A_1|^2 & = & \frac{\eta_2 }{\eta_1+\eta_2+\eta_1 \eta_2}\frac{\gamma (\vert F \vert -
1)}{4K_0^2}
\nonumber \\
|A_2|^2 & = & \frac{\eta_1 }{\eta_1+\eta_2+\eta_1
\eta_2}\frac{\gamma (\vert F \vert - 1)}{4K_0^2} \label{weak}
\end{eqnarray}
These expressions are still valid when $\Delta'$ and $c'$ are
different from zero, with $\eta_{1,2}$ given by:

\begin{equation}
\eta_{1,2}= \frac 12 \times\frac{1+c'^2+(\Delta'+4 \alpha' \, q_{1,2}^2)^2}
{[1+(c'+\Delta'+4 \alpha' \, q_{1,2}^2)^2] \, [1+(c'-\Delta'-4 \alpha'   \,
q_{1,2}^2)^2]}
\end{equation}
The asymptotic states of the dynamics, eqs. (\ref{weak}),
correspond to patterns built on two wavevectors $\vec q_1$ and
$\vec q_2$, as confirmed by the numerical results of the previous
section. However, when the kinetic coefficients of $x$ and $y$
field components are slightly different ($\Theta_x \neq
\Theta_y,|\Theta_x -\Theta_y| \ll \Theta_x +\Theta_y$) the growth
rates $\lambda_{1,2}$ of critical modes become different (see fig.
4) and the amplitude equations (\ref{finalmaster}) become:

\begin{eqnarray}
\partial_t A_{ 1} & = & \lambda_1 A_{ 1} -4 K_0^2 \, A_ 1
\,(|A_1|^2+|A_2|^2)
- 4\eta_1 K_0^2 \, A_1  \, |A_1|^2 \nonumber\\
\partial_t A_{2} & = & \lambda_2 A_{2} -4 K_0^2 \, A_ 2 \, (|A_1|^2+|A_2|^2)
- 4\eta_2 K_0^2 \, A_2 \, |A_2|^2
\label{lambda}
\end{eqnarray}
where $\lambda_1>\lambda_2$. In this case, patterns with
wavevector $\vec q_1$ only develop at, and slightly beyond
threshold. The corresponding steady state is

\begin{equation}
|A_1|^2=\frac{\lambda_1 }{4K_0^2(1+\eta_1)}, \,\,\,|A_2|=0
\end{equation}
Such patterns are stable provided $\lambda_2 <
\lambda_1/(1+\eta_1)$. When $\lambda_1$ reaches $\lambda_2
(1+\eta_1)$ they become unstable and in the domain where
$\lambda_1$ is larger than $\lambda_2 (1+\eta_1)$ patterns with
both wavevectors $\vec q_1$ and $\vec q_2$ are stable. Their
amplitude is given by:

\begin{eqnarray}
|A_1|^2 & = & \frac{\lambda_1 \,
(1+\eta_2)-\lambda_2}{4K_0^2(\eta_1+\eta_2+\eta_1 \eta_2)}
\nonumber \\
|A_2|^2 & = & \frac{\lambda_2 \,
(1+\eta_1)-\lambda_1}{4K_0^2(\eta_1+\eta_2+\eta_1 \eta_2)}
\end{eqnarray}

The two types of behavior (close and far from threshold) have been
described in our numerical results in the previous section.
Furthermore, on increasing pumping beyond threshold, one observes
a crossover between monomode patterns, with wavevectors
corresponding to the maximum growth rate, and bimodal ones. For
growth rates given by eqs.(\ref{eigen0}), transition  between mono
and bimodal patterns occurs, for small anisotropies ($\vert{
\Delta_x - \Delta_y \over \Delta_x + \Delta_y }\vert <<1$,$\vert {
\alpha_x - \alpha_y \over \alpha_x + \alpha_y}\vert <<1$) at:

\begin{equation}
|F|= 1 + ({ 2+\eta_2 \over 2\eta_2}){ \vert \alpha_x\Delta_x -
\alpha_y\Delta_y \vert \over \sqrt{\alpha_x\alpha_y }}
\end{equation}

\section{CONCLUSION}

In conclusion we have shown that standing wave intensity  patterns
can be generated in type-II optical parametric oscillators. They
appear spontaneously in the transverse plane when there is a
direct polarization coupling between the signal and idler fields,
produced for example by an intracavity quarter wave-plate. Such a
coupling gives also rise to two competing wavelengths in the
system.

In the transient dynamical regime after the pump is switched-on
above its threshold value, there is a competition between two
rings of unstable modes. This gives rise to transverse patterns
with different wavelengths for the real and imaginary part of the
FH's fields. We have described two dynamical regimes. In the first
one, which correspond to symmetric FH's parameters, the far field
is composed by two concentric rings. Real and imaginary part of
each of the FH's fields shows patterns with different wavelength.
For asymmetric FH's coefficients, the dynamical regime depends on
the distance to threshold. Near threshold, the dynamical process
of pattern competition leads to the dominance of a unique
wavelength, selecting a transverse stripe intensity pattern with
the same wavelength for signal and idler. Far from threshold, the
dynamics is equivalent to the symmetric case, so that the strength
of the external pump can be used to stabilize the two competing
wavelengths. Amplitude equations for the vectorial (critical)
modes have been derived and they confirm our numerical
observations. In particular, it is worth noting that the structure
of these equations reflect the vectorial nature of the fields,
which introduces nontrivial couplings between the modes.

\section{Acknowledgements}

We acknowledge financial support from the European Commission
projects QSTRUCT (FMRX-CT96-0077) and QUANTIM (IST-2000-26019) and
from the Spanish MCyT project BFM2000-1108. We thank Marco
Santagiustina for very helpful discussions on this topic.

\section{APPENDIX: Uniform phase-locked solutions}

In addition to the trivial uniform solution given by
eq.(\ref{trivial}), eqs.(\ref{master}) have  other uniform
stationary solutions for $\Delta_x / \Delta_y >0$ considered in
ref.\cite{Fabre00}. These are the dominant solutions when
transverse effects are not taken into account. For the sake of
completeness, a brief description of these solutions for
$\Delta_{x,y}<0$ is given here. Eqs. (\ref{master}) admit for
$A_{x,y}=a_{x,y} \, \exp(i \phi_{x,y})$ two uniform stationary
solutions, which take the form:
\begin{eqnarray}
\cos(\phi_d+\xi) & = & \frac{1-\Gamma^2}{2 \, |c| \, \Gamma} \nonumber \\
\cos(\phi_s) & = &  \frac{(\Delta_x + \Gamma \, \Delta_y)-2 \, |c|
\, \Gamma
\, \sin(\phi_d+\xi)}{2 \, K_0 \, E_0 \, c_p} \nonumber \\
a_x^2 & = & \frac{2 \, c_p \, E_0 \, \Gamma \,
\sin(\phi_s)-1-\Gamma^2}{4
\, c_p \, \Gamma^2} \nonumber \\
a_y & = & \Gamma \, a_x \label{locked}
\end{eqnarray}
where $ \Gamma^2=\Delta_x / \Delta_y$, $c_p=1/(1-|c'|^2)$ and
$\phi_{d,s}= \phi_x \mp \phi_y$. The existence of these solutions
requires for $\Gamma \neq 1$ that $|c|
>|1-\Gamma^2|/(2 \, \Gamma)$. This relation defines a circle in the
complex plane of $c$ inside which no stationary uniform solutions
exist.

The first of eqs.(\ref{locked}) indicates that these homogeneous
solutions are self phase-locked. Due to the fact that the function
 ${\it arccos}$ is a multi-valued function in the range
$[-\pi,\pi]$, two branches of uniform phase-locked solutions
exist. Each branch has a different threshold. Uniform solutions
are never spontaneously observed for $\Delta_{x,y}<0$ because they
have a larger threshold of instability than the solution with a
finite wavenumber. The threshold of the uniform branch with lower
threshold is indicated by a diamond in figure 1. Note that this is
larger than the threshold for pattern formation $|F_c|=1$. The
second branch of uniform solutions, with a larger threshold value,
is unstable and it is not observed, not even for positive FH's
detunings where uniform solutions dominate and domain walls
between them have been reported \cite{Nos00,Nos01}. The
homogeneous solutions described here are different of those in
Fig. 10 obtained in the limit of $q_1=0$.

\begin{figure}

\centerline{\psfig{figure=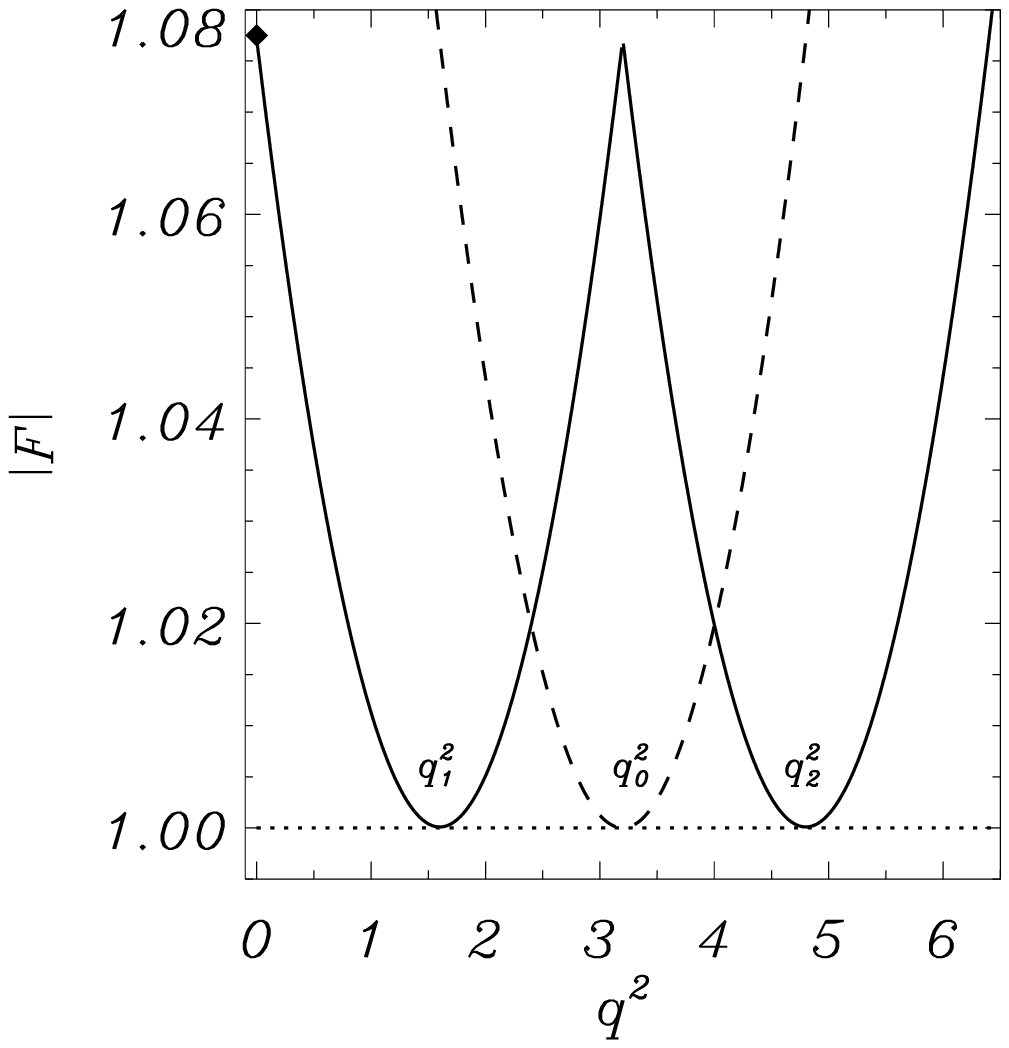,width=5in}}

\caption{Scaled threshold of instability $F$ for the trivial
stationary solution (eq.(\ref{trivial})) as a function of $q^2$
for the case of symmetric FH's coefficients. Solid (dashed) curve
gives the threshold value of $F$ for $c \neq 0$ ($c=0$). The
critical threshold $|F_c|=1$ is  indicated as a dotted line. The
instability takes place  at two different wavectors indicated by
$q_1$ and $q_2$ ($q_0$ denotes the unstable wavector for $c=0$).
Parameters are $K_0=1$, $\Delta_{x,y}=-0.8$, $\Delta_{x,y}'=0$,
$\alpha_{x,y}=0.25$,   $\alpha_{x,y}'=0.125$, $c=0.4$ and
$c'=0.01$. The diamond indicates the threshold of instability for
homogeneous perturbations ($\vec q=0$).}

\end{figure}

\newpage

\begin{figure}

\centerline{\psfig{figure=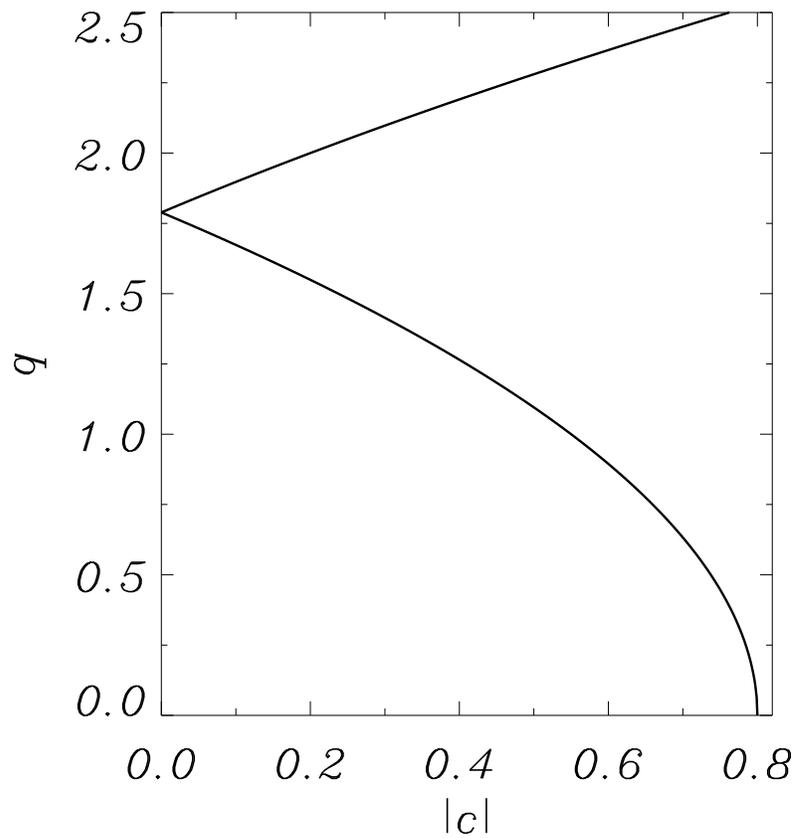,width=5in}}

\caption{Wavenumber of the critical unstable modes as a function
of $|c|$. The lower branch corresponds to $q_1$ and the upper one
to $q_2$. Parameter values as in figure 1. }

\end{figure}

\newpage

\begin{figure}

\centerline{\psfig{figure=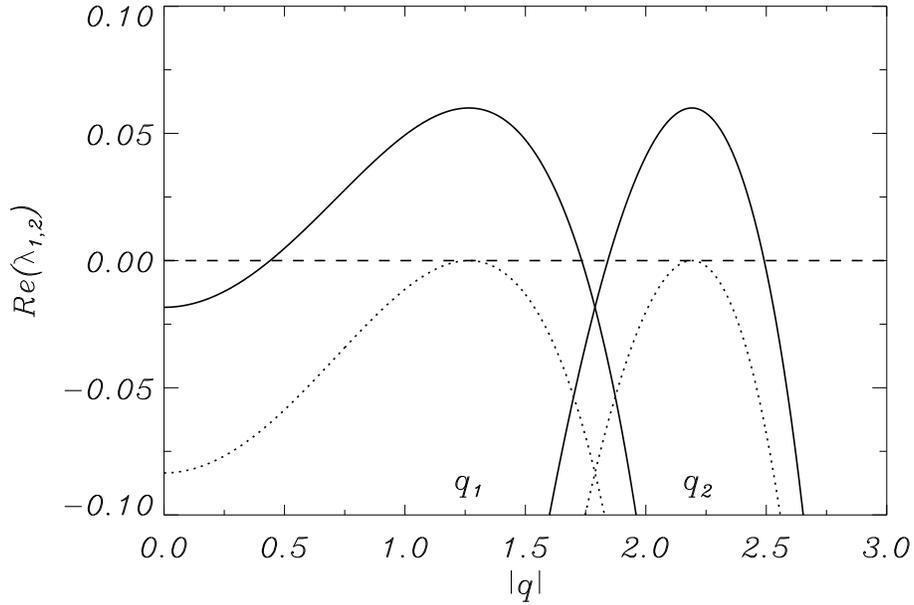,width=5in}}

\caption{Real part of the eigenvalues $ \lambda_{1,2}$ as a
function of $|\vec q|$ for the case of symmetric FH's
coefficients. The solid lines correspond to $F=1.06$ and the
dotted to criticality, $F=F_c=1$. Left branches correspond to
$\lambda_1$ and the right ones to $\lambda_2$. Here
$\gamma_{x,y}=\gamma_{x,y}'=1$ and the values of the other
parameters  are the same as in Fig. 1. The most unstable modes
$q_1$ and $q_2$ correspond to the maximum of each line. The level
$Re(\lambda_{1,2})=0$ is also indicated as a dashed horizontal
line as a reference.}

\end{figure}

\newpage

\begin{figure}

\centerline{\psfig{figure=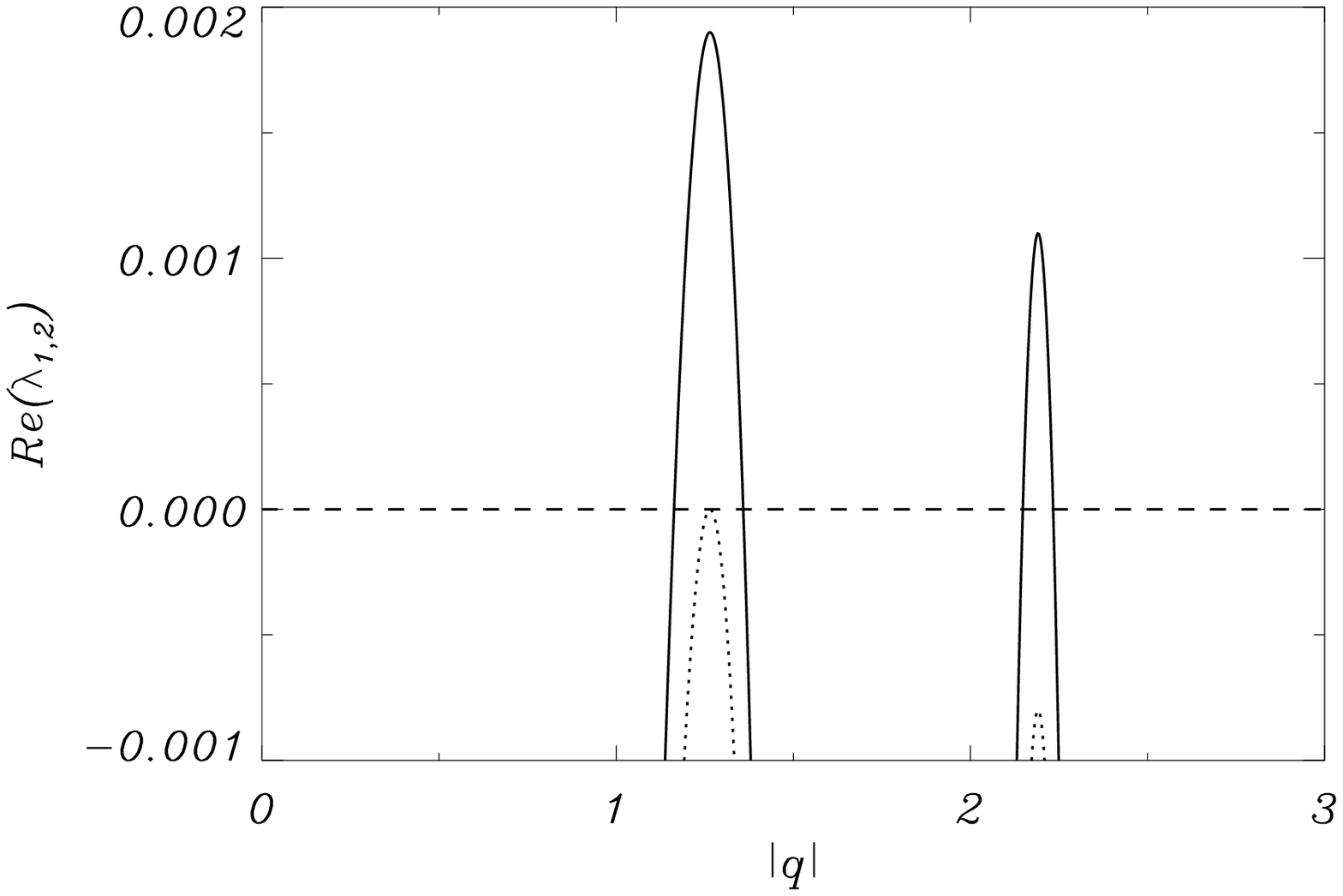,width=5in}}

\caption{Growth rate of the unstable modes as a function of $|\vec
q|$ in a case in which  the FH's coefficients are different. Fig.
4.a) Near threshold: $F=1.0019$ (solid line) and $F=1$ (dotted
line). Fig. 4.b) Far from threshold: $F=1.02$. Coefficients values
are $K_0=1$, $\Delta_{x,y}=-0.8$, $\Delta_{x,y}'=0$,
$\gamma_{x}=\gamma_{x}'=0.9901$, $\gamma_{y}=\gamma_{y}'=1.01$,
$\alpha_{x}=0.2475$, $\alpha_{y}=0.2525$, $\alpha_{x,y}'=0.125$,
$c=0.4$ and $c'=0.01$.}

\centerline{\psfig{figure=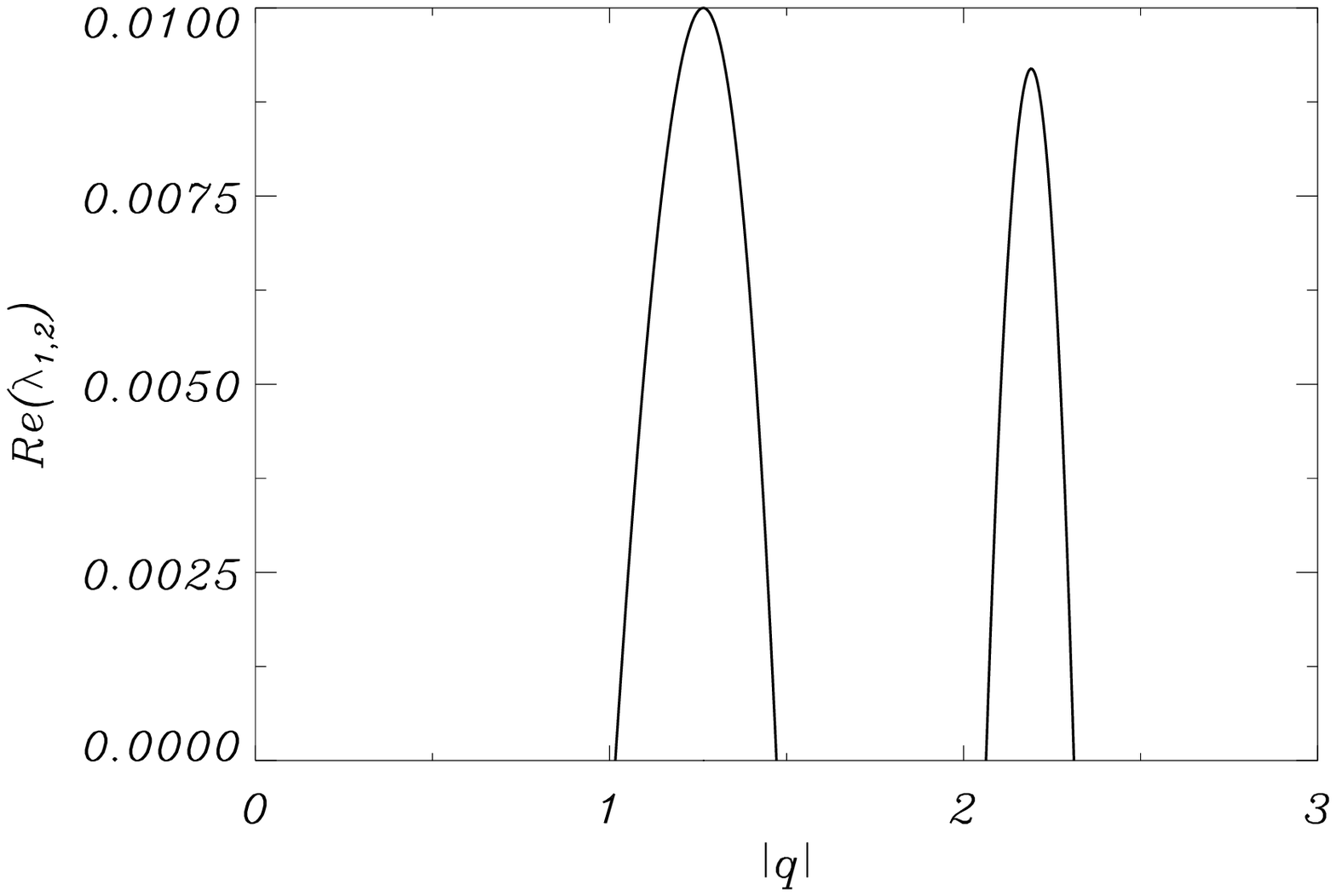,width=5in}}

\centerline{Figure 4.b}

\end{figure}

\newpage

\begin{figure}

\centerline{\psfig{figure=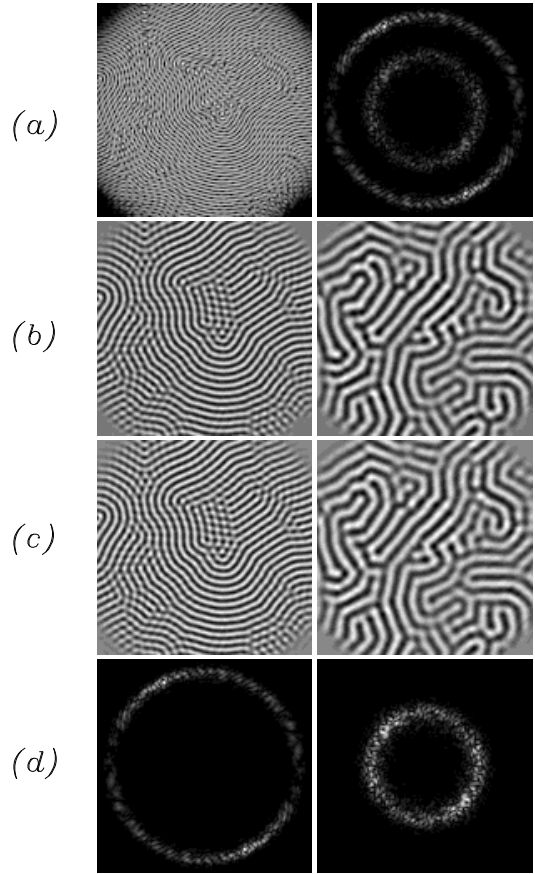,width=5in}}

\vspace{5cm}

\caption{A snapshot at $t=900$ of the FH's fields spontaneously
generated from random initial conditions close to the trivial
steady-sate given by (3).  a) left: near field $|A_x|$; right: far
field $|\widetilde{A}_{\vec q}|$. b) $A_x$ field. Left: $Re(A_x)$;
right: $Im(A_x)$. c) $A_y$ field. Left: $Im(A_y)$; right:
$Re(A_y)$. d) Absolute value of the Fourier transform. Left:
$Re(A_x)$; right: $Im(A_x)$. Parameters are $F=1.16$, $K_0=1$,
$\Delta_{x,y}=-0.8$, $\Delta_{x,y}'=0$,
$\gamma_{x,y}=\gamma_{x,y}'=1$, $\alpha_{x,y}=0.25$,
$\alpha_{x,y}'=0.125$, $c=0.4$ and $c'=0.01$.}

\end{figure}

\newpage

\begin{figure}

\centerline{\psfig{figure=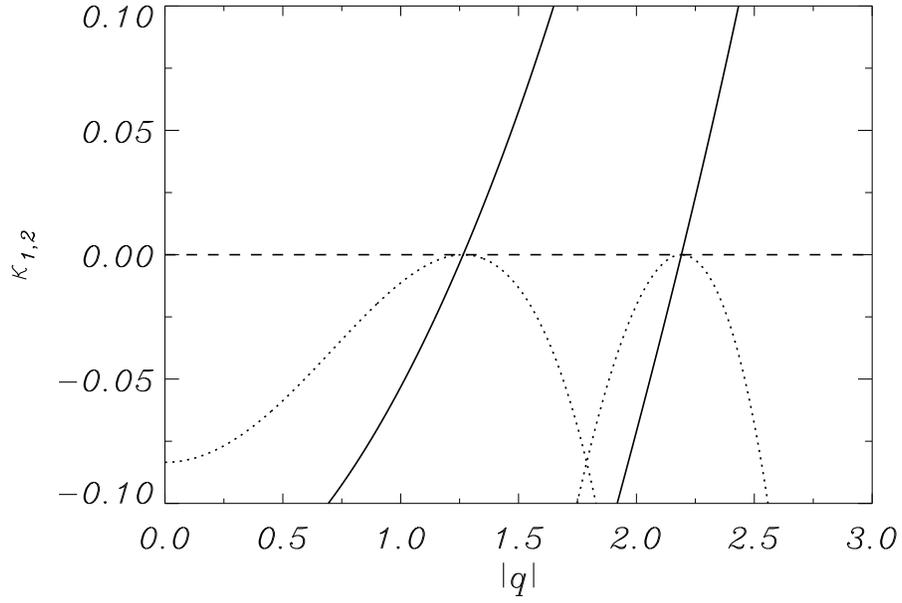,width=5in}}

\caption{Amplitudes $\kappa_{1,2}$ (solid line) as a function of
$|\vec q|$. The real part of the critical eigenvalues
$\lambda_{1,2}$ (dotted lines) and the zero level (dashed line)
are also plotted as reference. Here $F=1$ and the values of the
other parameters are the same of figure 3.}

\end{figure}

\newpage

\begin{figure}

\centerline{\psfig{figure=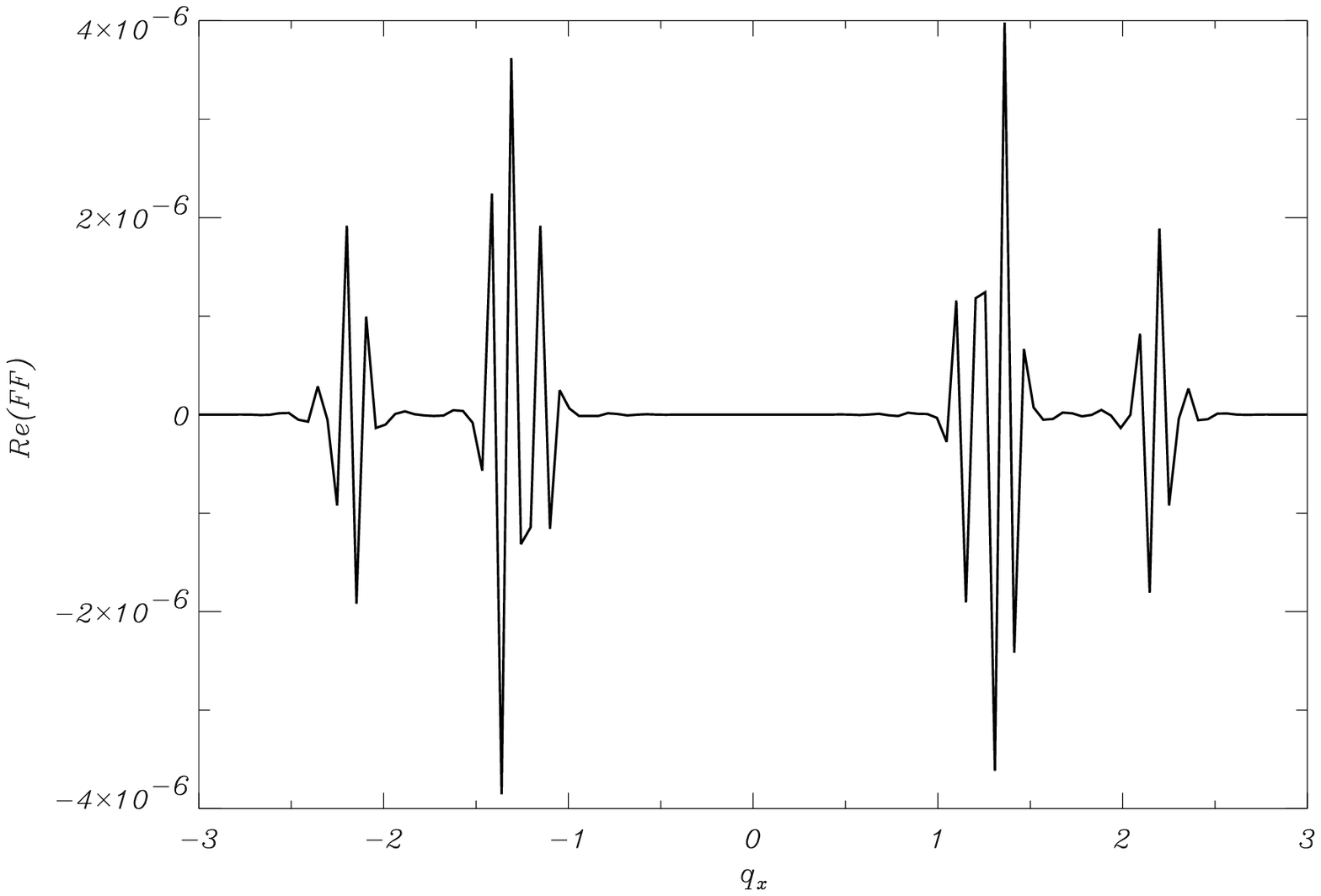,width=5in}}

\caption{A typical result for a cut of the far field ($FF$)
amplitude $\widetilde{A}_{\vec q} (t)$ along the line $q_y=0$ for
the $A_x$ field  during the transient state. a) Real part; b)
Imaginary part. Parameters are $F=1.06$, $K_0=1$,
$\Delta_{x,y}=-0.8$, $\Delta_{x,y}'=0$,
$\gamma_{x}=\gamma_{x}'=0.9901$, $\gamma_{y}=\gamma_{y}'=1.01$,
$\alpha_{x}=0.2475$, $\alpha_{x}=0.2525$, $\alpha_{x,y}'=0.125$,
$c=0.4$ and $c'=0.01$.}

\centerline{\psfig{figure=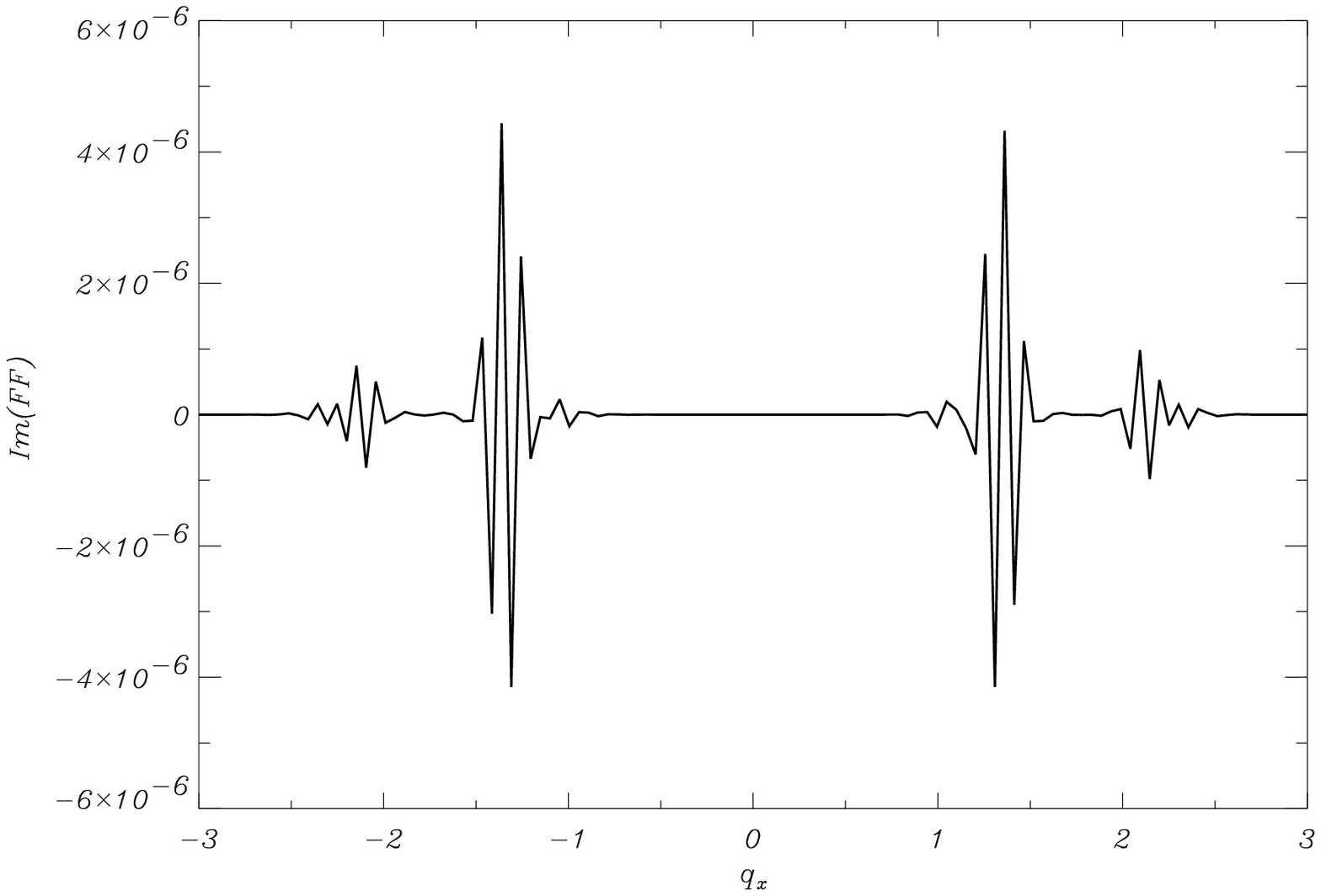,width=5in}}

\centerline{Figure 7.b}

\end{figure}

\newpage

\begin{figure}

\caption{Snapshots as in Fig. 5 at time $t=7800$. Parameters are
the same of Fig.5}

\centerline{\psfig{figure=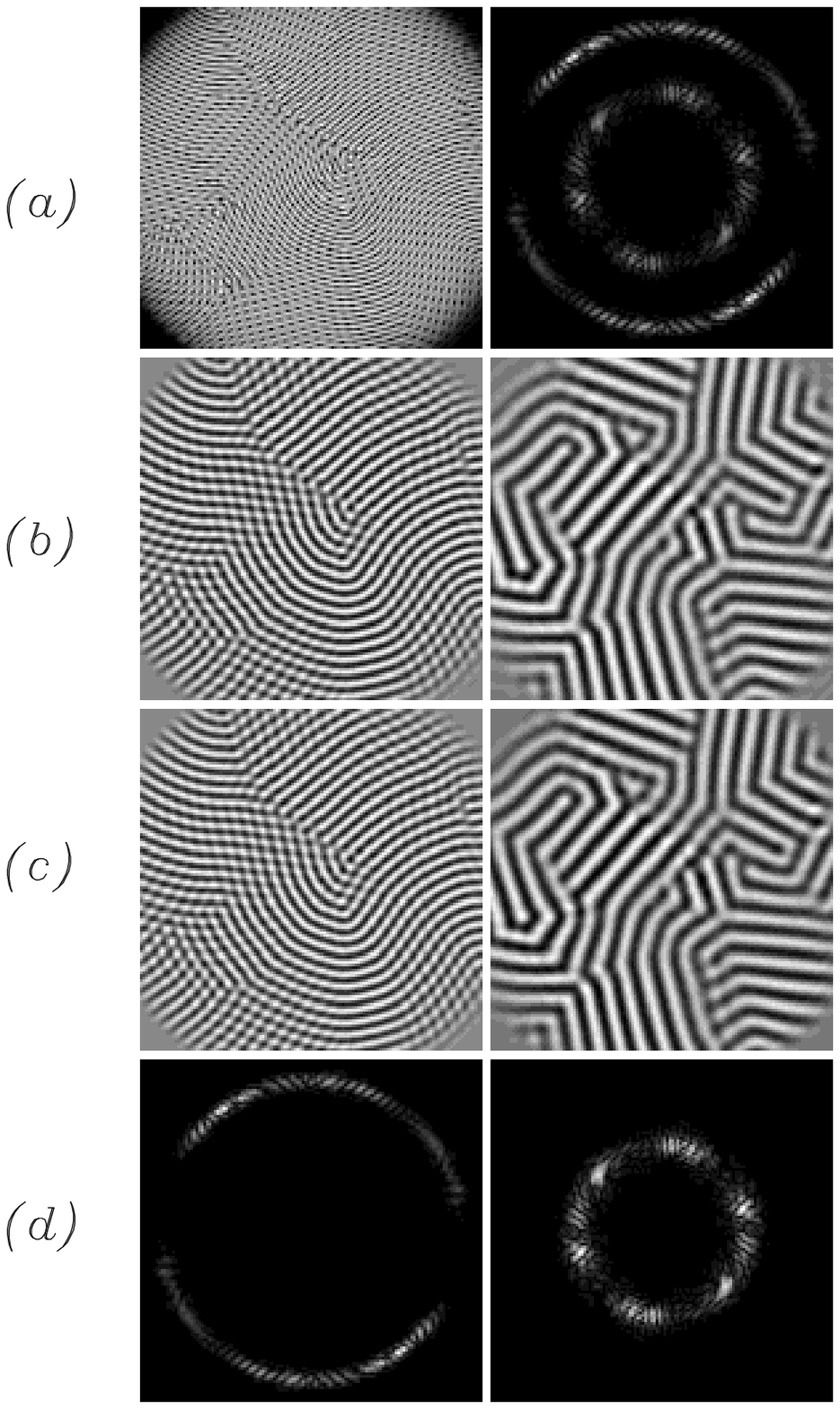,width=5in}}

\end{figure}

\newpage

\begin{figure}

\centerline{\psfig{figure=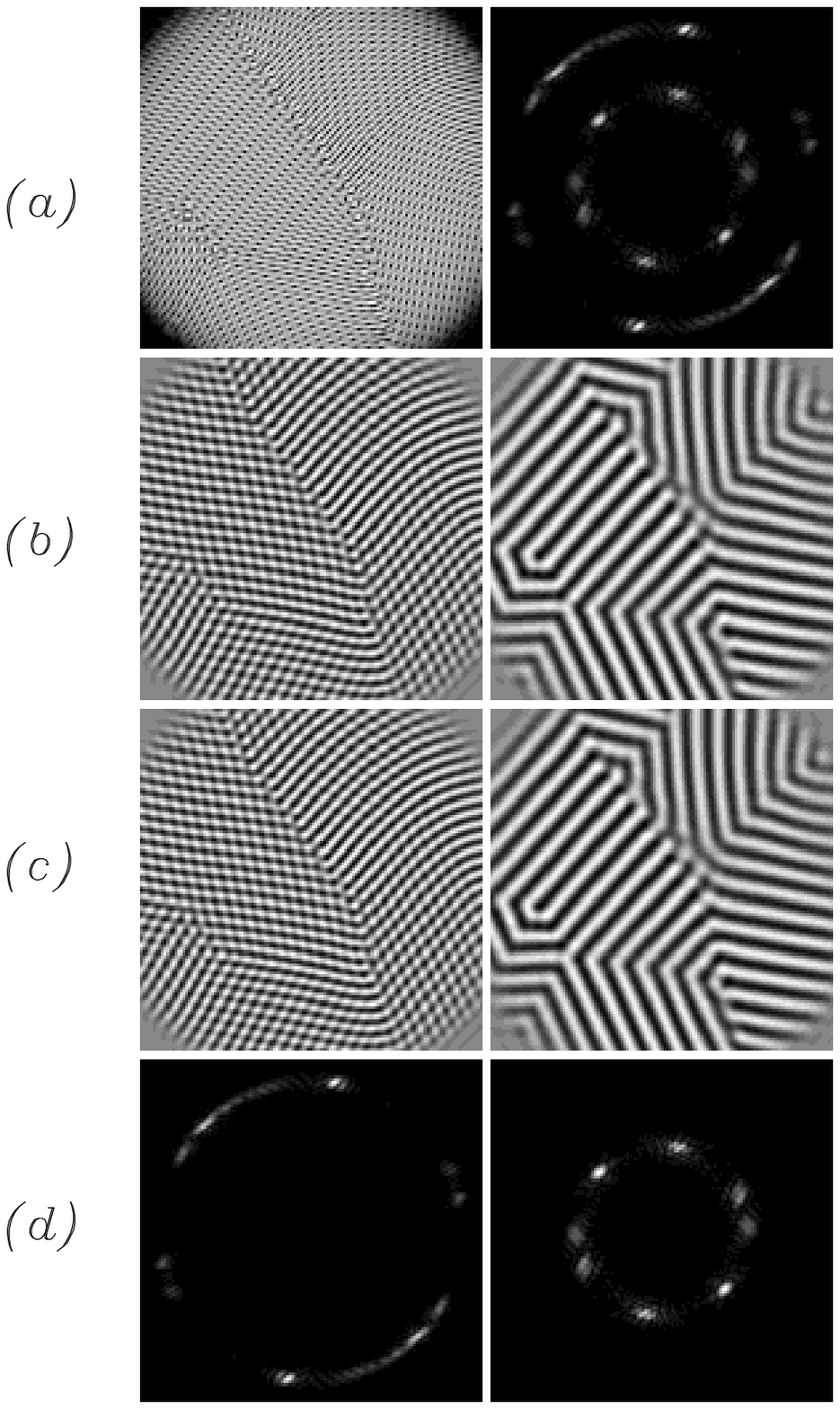,width=5in}}

\vspace{5cm}

\caption{Snapshots as in Fig. 5 at time $t=42800$. Parameters are
the same of Fig.5 }

\end{figure}

\newpage

\begin{figure}

\centerline{\psfig{figure=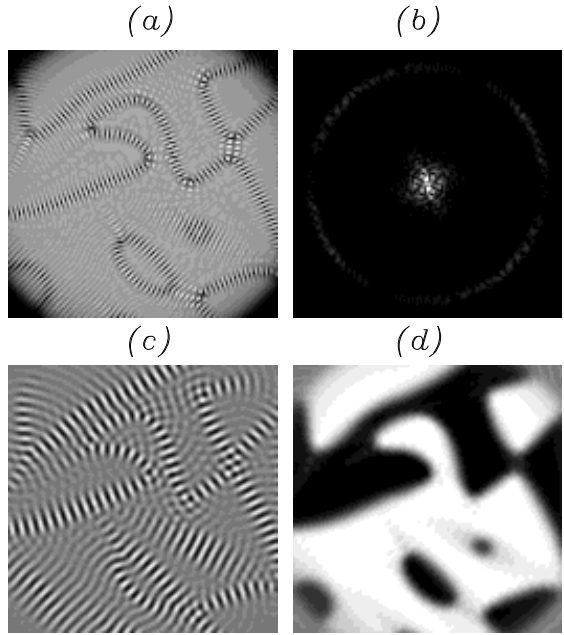,width=5in}}

\caption{A snapshot at time $t=3000$  of the field $A_x$
spontaneously generated from random initial conditions for a case
in which the mode $\vec q=0$ is one of the most unstable modes.
Figures a), b), c) and d) show respectively the intensity, far
field, real part and imaginary part of the signal field. In c) and
d) patterns with  very different wavelength can be appreciated.
Parameter values are $F=1.08$, $K_0=1$, $\Delta_{x,y}=-0.6$,
$\Delta_{x,y}'=0$,$\gamma_{x}=\gamma_{x}'=\gamma_{y}=\gamma_{y}'=1$,
$\alpha_{x}=\alpha_{y}=0.25$, $c=0.6$ and $c'=0.01$.}

\end{figure}

\newpage

\begin{figure}

\caption{Snapshots as in Fig. 5 at time $t=100$. Parameter values
are  $E_0=1.002$, $K_0=1$, $\Delta_{x,y}=-0.8$, $\Delta_{x,y}'=0$,
$\gamma_{x}=\gamma_{x}'=0.9901$, $\gamma_{y}=\gamma_{y}'=1.01$,
$\alpha_{x}=0.2475$, $\alpha_{y}=0.2525$, $\alpha_{x,y}'=0.125$,
$c=0.4$ and $c'=0.01$.}

\centerline{\psfig{figure=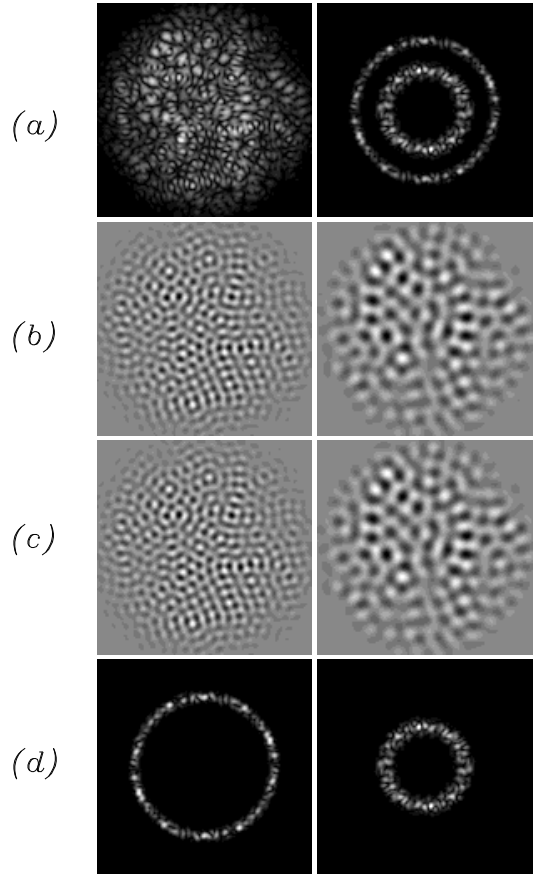,width=5in}}

\end{figure}

\newpage

\begin{figure}

\caption{Snapshots as in Fig. 11  at time $t=3200$.}

\centerline{\psfig{figure=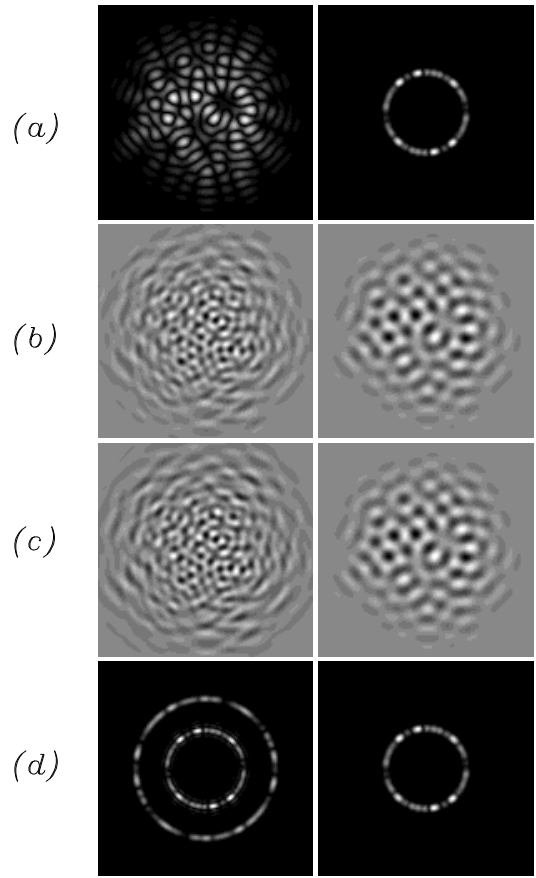,width=5in}}

\end{figure}

\newpage

\begin{figure}

\centerline{\psfig{figure=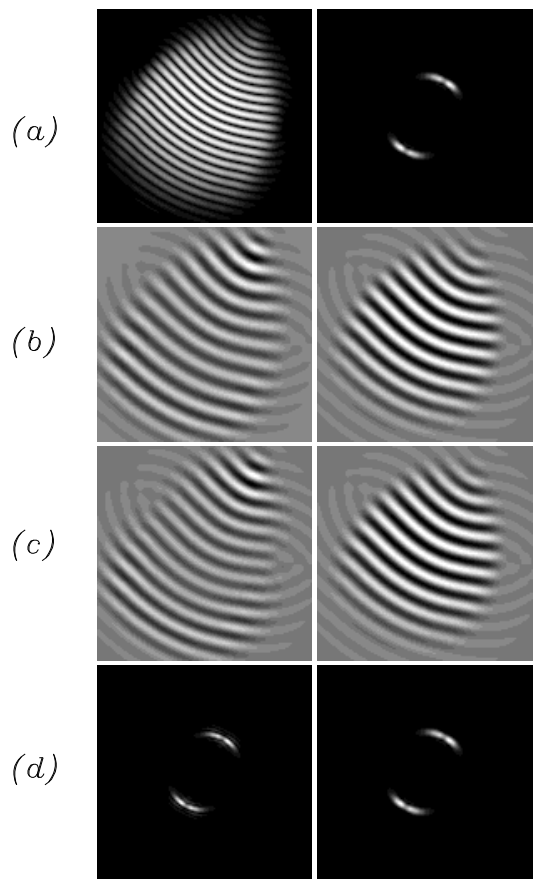,width=5in}} \vspace{5cm}

\caption{Snapshots as in Fig. 11 at time $t=17600$.}

\end{figure}

\newpage

\begin{figure}

\centerline{\psfig{figure=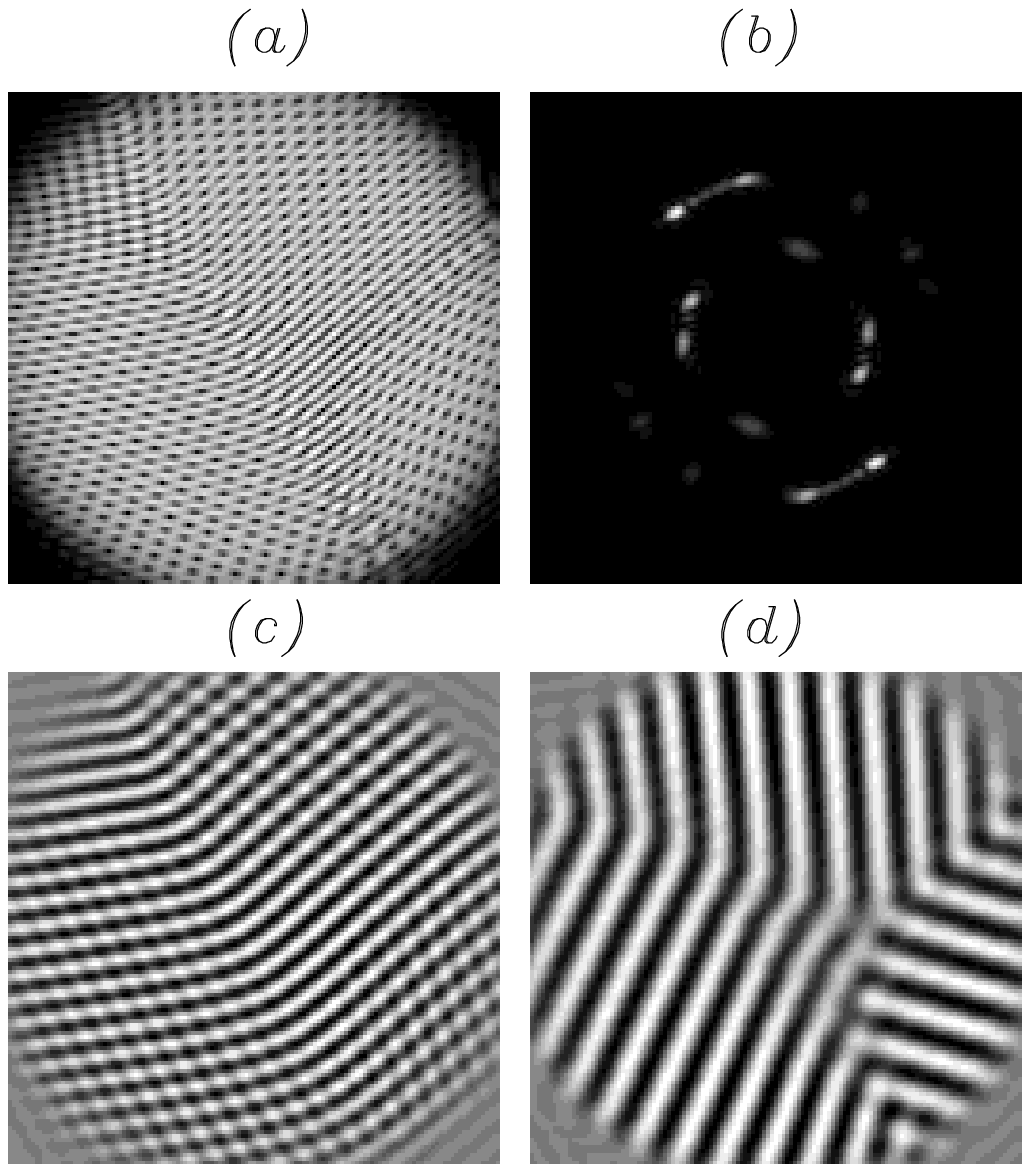,width=5in}}

\caption{A snapshot at time $t=50500$ of the field $A_x$
spontaneously generated from random initial conditions close to
the trivial steady-state in the asymmetric case. a), b), c) and d)
show the intensity, far field, real  part and imaginary part of
$A_x$ respectively. Parameters are as in figure 11 except
$F=1.06$.}

\end{figure}


\begin{thebibliography}{99}
\bibitem{numerical} Eqs. (\ref{master}) have been integrated using the
algorithm described in  ref. \cite{pre}. We use a grid of
256$\times$256 samples with grid space $\Delta x=\Delta y=0.3125$
and time step $\Delta t=0.01$. For all cases we use a flat-top
super-Gaussian pump beam $E_0(x,y)$.
\end{thebibliography}
\end{document}